\documentclass[12pt]{article}
\usepackage{amsfonts}

\begin{document}

\title{Beyond Octonions}
\author{Khaled Abdel-Khalek \footnote{khaled@le.infn.it}\ \footnote{Address after 15 Feberuary 2000: 
Feza G\"ursey Institute, P.O. Box 6, 81220 \c{C}engelk\"oy, 
 \'Istanbul, Turkey}\\ Dipartimento di Fisica, Universit\`a di Lecce\\
- Lecce, 73100, Italy -
}
\date{February 2000}
\maketitle

\begin{abstract}
We investigate Clifford Algebras structure over non-ring division algebras. We show how 
projection over the real field produces the standard Attiyah-Bott-Shapiro classification.
\end{abstract}

Quaternions  and octonions may be presented as a linear algebra over the
field of real numbers $\Bbb{R}$ with a general element of the form 
\begin{equation}
Y=y_{0}e_{0}+y_{i}e_{i},\;\;\;\;\;\;\;y_{0},y_{i}\in \Bbb{R}  \label{gen}
\end{equation}
where $i=1,2,3$ for quaternions $\Bbb{H}$ and $i=1..7$ for octonions $\Bbb{O}
$. \ We always use Einstein's summation convention. The $e_{i}$ are
imaginary units, for quaternions 
\begin{eqnarray}
e_{i}e_{j} &=&-\delta _{ij}+\epsilon _{ijk}e_{k}, \\
e_{i}e_{0} &=&e_{0}e_{i}=e_{i}, \\
e_{0}e_{0} &=&e_{0},
\end{eqnarray}
where $\delta _{ij}$ is the Kronecker delta and $\epsilon _{ijk}$ is the
three dimensional Levi--Cevita tensor, as $e_{0}=1$ when there is no
confusion we omit it. Octonions have the same structure, only we must
replace $\epsilon _{ijk}$ by the octonionic structure constant $f_{ijk}$
which is completely antisymmetric and equal to one for any of the following
three cycles 
\begin{equation}
123,\;145,\;176,\;246,\;257,\;347,\;\;365.
\end{equation}
The important feature of real, complex, quaternions and octonions is the
existence of an inverse for any non-zero element. For the generic
quaternionic or octonionic element given in (\ref{gen}), we define the
conjugate $Y^{*}$ as an involution $\left( Y^{*}\right) ^{*}=Y$, such that 
\begin{equation}
Y^{*}=y_{0}e_{0}-y_{i}e_{i},
\end{equation}
introducing the norm as $N\left( Y\right) \equiv \left\| Y\right\|
=YY^{*}=Y^{*}Y$ \ then the inverse is 
\begin{equation}
Y^{-1}=\frac{Y^{*}}{\left\| Y\right\| }.
\end{equation}
The Norm is nondegenerate and positively definite. We have the decomposition
property 
\begin{equation}
\left\| XY\right\| =\left\| X\right\| \;\;\left\| Y\right\|  \label{dec}
\end{equation}
$N\left( xy\right) $ being nondegenerate \ and positive definite obeys the
axioms of the scalar product.

Going to higher dimensions, we define ``hexagonions'' ($\Bbb{X}$) by
introducing a new element $e_{8}$ such that 
\begin{equation}
\begin{array}{llll}
\Bbb{X} & = & \Bbb{O}_{1}+\Bbb{O}_{2}e_{8} &  \\ 
& = & x_{0}e_{0}+\ldots +x_{16}e_{16}. & x_{\mu }\in \Bbb{R}
\end{array}
\label{ffff}
\end{equation}
and 
\begin{equation}
e_{i}e_{j}=-\delta _{ij}+C_{ijk}e_{k}.
\end{equation}
Now, we have to find a suitable form of the completely antisymmetric tensor $%
C_{ijk}$. Recalling how the structure constant is written for octonions 
\begin{eqnarray}
\Bbb{O} &=&\Bbb{Q}_{1}+\Bbb{Q}_{2}e_{4}  \nonumber \\
&=&x_{0}e_{0}+\ldots +x_{7}e_{7},  \label{a4}
\end{eqnarray}
where $\Bbb{Q}$ are quaternions, we have already chosen the convention $%
e_{1}e_{2}=e_{3}$ which is extendable to (\ref{a4}). We set $%
e_{1}e_{4}=e_{5} $, $e_{2}e_{4}=e_{6}$ and $e_{3}e_{4}=e_{7}$, but we still
lack the relationships between the remaining possible triplets, $%
\{e_{1},e_{6},e_{7}\};$ $\{e_{2},e_{5},e_{7}\};$ $\{e_{3},e_{5},e_{6}\}$
which can be fixed by using 
\[
\begin{array}{c}
e_{1}e_{6}=e_{1}(e_{2}e_{4})=-(e_{1}e_{2})e_{4}=-e_{3}e_{4}=-e_{7}, \\ 
e_{2}e_{5}=e_{2}(e_{1}e_{4})=-(e_{2}e_{1})e_{4}=+e_{3}e_{4}=+e_{7}, \\ 
e_{3}e_{5}=e_{3}(e_{1}e_{4})=-(e_{3}e_{1})e_{4}=-e_{2}e_{4}=-e_{6}.
\end{array}
\]
These cycles define all the
structure constants for octonions. Returning to $\Bbb{X}$, we have the seven
octonionic conditions, and the decomposition (\ref{ffff}). We set $
e_{1}e_{8}=e_{9},\;e_{2}e_{8}=e_{A},\;e_{3}e_{8}=e_{B},\;e_{4}e_{8}=e_{C},%
\;e_{5}e_{8}=e_{D},\;e_{6}e_{8}=e_{E},\;e_{7}e_{8}=e_{F}$ where $%
A=10,\;B=11,\;C=12,\;D=13,\;E=14$ and $F=15$. The other elements of the
multiplication table may be chosen in analogy with (\ref{a4}). Explicitly,
the 35 hexagonionic triplets are 
\[
\begin{array}{ccccccc}
(123), & (145), & (246), & (347), & (257), & (176), & (365), \\ 
(189), & (28A), & (38B), & (48C), & (58D), & (68E), & (78F), \\ 
(1BA), & (1DC), & (1EF), & (29B), & (2EC), & (2FD), & (3A9), \\ 
(49D), & (4AE), & (4BF), & (3FC), & (3DE), & (5C9), & (5AF), \\ 
(5EB), & (6FD), & (6CA), & (6BD), & (79E), & (7DA), & (7CB).
\end{array}
\]
This can be extended for any generic higher dimensional $\Bbb{F}^{n}$.

It can be shown by using some combinatorics that the number of such triplets 
$N$ for a general $\Bbb{F}^{n}$ algebra is ($n>1$) 
\begin{equation}
N={\frac{~~~\left( 2^{n}-1\right) !~~~}{~\left( 2^{n}-3\right) !~~~3!~}},
\end{equation}
giving 
\[
\begin{array}{cccc}
\Bbb{F}^{n} & n & ~~~~dim~~~~ & N \\ 
\Bbb{Q} & 2 & 4 & 1 \\ 
\Bbb{O} & 3 & 8 & 7 \\ 
\Bbb{X} & 4 & 16 & 35 \\ 
&  & and\ so\ on. & 
\end{array}
\]
One may notice that for any non-ring division algebra $\left( \Bbb{F},\
n>3\right) $,\ $N>dim(\Bbb{F}^{n})$ except when dim = $\infty ,$ i.e. a
functional Hilbert space with a Cliff(0,$\infty $) structure.

It is clear that for any ring or non--ring division algebras, $%
e_{i},e_{j}\in \Bbb{F}^{n}$, we have 
\begin{equation}
\left\{ e_{i},e_{j}\right\} =-2\delta _{ij}.
\end{equation}
As we explained in \cite{h1} and \cite{h2}, treating quaternions and
octonions as elements of $R^{4}$ and $R^{8}$ respectively, we can find the
full set of matrices $R\left( 4\right) $ and $R\left( 8\right) $ that
corresponds to any elements $e_{i}$ explicitly 
\begin{eqnarray}
&& 
\begin{array}{llll}
for\;quaternions & e_{i} & \longleftrightarrow & (\Bbb{E}_{i})_{\alpha \beta
}=\delta _{i\alpha }\delta _{\beta 0}-\delta _{i\beta }\delta _{\alpha
0}+\epsilon _{i\alpha \beta }, \\ 
\left\{ E_{i},E_{j}\right\} =-2\delta _{ij} &  & i,j=1..3, & \alpha ,\beta
=1..4, \\ 
for\;octonions & e_{i} & \longleftrightarrow & (\Bbb{E}_{i})_{\alpha \beta
}=\delta _{i\alpha }\delta _{\beta 0}-\delta _{i\beta }\delta _{\alpha
0}+f_{i\alpha \beta }, \\ 
\left\{ E_{i},E_{j}\right\} =-2\delta _{ij} &  & i,j=1..7, & \alpha ,\beta
=1..8
\end{array}
\nonumber \\
&&  \label{eee}
\end{eqnarray}

Following, the same translation idea projecting our algebra $\Bbb{X}$ over $%
\Bbb{R}^{16}$, any $\Bbb{E}_{i}$ is given by a relation similar to that
given in (\ref{eee}), 
\begin{equation}
(\Bbb{E}_{i})_{\alpha \beta }=\delta _{i\alpha }\delta _{\beta 0}-\delta
_{i\beta }\delta _{\alpha 0}+C_{i\alpha \beta }.
\end{equation}
But contrary to quaternions and octonions, the Clifford algebra (over the
real field $R^{16}$)closes only for a subset of these $E_{i}$'s, namely 
\begin{equation}
\{\Bbb{E}_{i},\Bbb{E}_{j}\}=-2\delta _{ij}\quad \mbox{for}\quad
i,j,k=1\ldots 8\;\;not\;1...15.  \label{nrd}
\end{equation}
Because we have lost the ring division structure. We can find easily that
another ninth $\Bbb{E}_{i}$ \ can be constructed, in agreement with the
Clifford algebra classification \cite{h3}. There is no standard$\footnote{%
Look to \cite{h2} for a non standard representation.}$ 16 dimensional
representation for $Cliff\left( 15\right) $. Following this procedure, we
can give a simple way to write real Clifford algebras over any arbitrary
Euclidean dimensions.

Sometimes, a specific multiplication table may be favored. For example in
soliton theory, the existence of a symplectic structure related to the
bihamiltonian formulation of integrable models is welcome. It is known from
the Darboux theorem, that locally a symplectic structure is given up to a
minus sign by 
\begin{equation}
\mathcal{J}_{dim\times dim}=\left( 
\begin{array}{cc}
0 & -\mathbf{1}_{\frac{dim}{2}} \\ 
\mathbf{1}_{\frac{dim}{2}} & 0
\end{array}
\right) ,
\end{equation}
this fixes the following structure constants 
\begin{eqnarray}
&&C_{\left( {\frac{dim}{2}}\right) 1\left( {\frac{dim}{2}}+1\right) }=-1, \\
&&C_{\left( {\frac{dim}{2}}\right) 2\left( {\frac{dim}{2}}+2\right) }=-1, \\
&&~~~~~~~\vdots \\
&&C_{\left( {\frac{dim}{2}}\right) \left( {\frac{dim}{2}}-1\right) \left(
dim-1\right) }=-1,
\end{eqnarray}
which is the decomposition that we have chosen in (\ref{a4}) for octonions 
\begin{equation}
C_{415}=C_{426}=C_{437}=-1.
\end{equation}
Generally our symplectic structure is 
\begin{equation}
\left( 1|\Bbb{E}_{\left( {\frac{dim}{2}}\right) }\right) _{\alpha \beta
}=\delta _{0\alpha }\delta _{\beta \left( {\frac{dim}{2}}\right) }-\delta
_{0\beta }\delta _{\alpha \left( {\frac{dim}{2}}\right) }-\epsilon _{\alpha
\beta \left( {\frac{dim}{2}}\right) }.
\end{equation}
Moreover some other choices may exhibit a relation with number theory and
Galois fields \cite{h4}. It is highly non-trivial how Clifford algebraic
language can be used to unify many distinct mathematical notions such as
Grassmanian \cite{h5}, complex, quaternionic and symplectic structures.

The main result of this section, the non-existence of standard associative
16 dimensional representation of $Cliff\left( 0,15\right) $ is in agreement
with the Atiyah--Bott--Shapiro classification of real Clifford algebras 
\cite{h3}. In this context, the importance of ring division algebras can also be
deduced from the Bott periodicity \cite{h6}.

I would like to thank P. Rotelli for some useful comments.


\newpage

\begin{thebibliography}{9}
\bibitem{h1}  S.~De~Leo~and~K.~Abdel-Khalek, J. Math. Phys., 38 (1997) 582.
\bibitem{h2}  K.~Abdel-Khalek, Int. J. Mod. Phys. A13 (1998) 223.
\bibitem{h3}  M.~F.~Atiyah, R.~Bott and A.~Shapiro, Topology \textbf{3}
(Suppl. 1) (1964) 3.
\bibitem{h4}  G.~Dixon, BRX TH-372, hep-th/9503053.
\bibitem{h5}  K.~Abdel-Khalek, Int. J. Mod. Phys. A13 (1998) 569.
\bibitem{h6}  K.~B.~Marathe and G.~Martucci, The mathematical
foundations of gauge theories, Amsterdam, North-Holland, 1992.
\end{thebibliography}
\end{document}